\documentclass[sn-mathphys,Numbered]{sn-jnl}

\usepackage{graphicx}%
\usepackage{multirow}%
\usepackage{amsmath,amssymb,amsfonts}%
\usepackage{amsthm}%
\usepackage{mathrsfs}%
\usepackage[title]{appendix}%
\usepackage{xcolor}%
\usepackage{textcomp}%
\usepackage{manyfoot}%
\usepackage{booktabs}%
\usepackage{algorithm}%
\usepackage{algorithmicx}%
\usepackage{algpseudocode}%
\usepackage{listings}%

\usepackage{comment}
\usepackage{caption}
\usepackage{subcaption}

\begin{document}

\title[Mass Ratio Dependence of Three-Body Resonance Lifetimes in 1D and 3D]{Mass Ratio Dependence of Three-Body Resonance Lifetimes in 1D and 3D}


\author*[1]{\fnm{Lucas} \sur{Happ}}\email{lucas.happ@riken.jp}

\author[1]{\fnm{Pascal} \sur{Naidon}}

\author[1,2]{\fnm{Emiko} \sur{Hiyama}}

\affil*[1]{\orgdiv{Few-body Systems in Physics Laboratory}, \orgname{RIKEN Nishina Center for Accelerator-Based Science}, \city{Wak\={o}}, \state{Saitama} \postcode{351-0198}, \country{Japan}}

\affil[2]{\orgdiv{Department of Physics}, \orgname{Tohoku University}, \city{Sendai}, \state{Miyagi} \postcode{980-8578}, \country{Japan}}

\abstract{We present a theoretical study of resonance lifetimes in a two-component three-body system, specifically examining the decay of three-body resonances into a deep dimer and an unbound particle. Utilising the Gaussian expansion method together with the complex scaling method, we obtain the widths of these resonances from first principles. We focus on mass ratios in the typical range for mixtures of ultracold atoms and reveal an intriguing dependence of the resonance widths on the mass ratio: as the mass ratio increases, the widths exhibit oscillations on top of an overall decreasing trend. In particular, for some mass ratios the resonance width vanishes, implying that the resonance becomes in fact stable. Notably, near the mass ratio for Caesium-Lithium mixtures, we obtain nearly vanishing widths of the resonances which validates to treat them in the bound-state approximation. In addition, we perform our analysis of the resonance widths in both one and three dimensions and find a qualitatively similar dependence on the mass ratio.}




\maketitle
\section{Introduction}\label{sec:intro}
Resonances are metastable quantum states that can spontaneously undergo a transition into a continuum state~\cite{moiseyev1998}. This fact makes them ubiquitous in few-body systems due to the presence of possibly many continua induced by the existence of breakup thresholds into two or more subsystems. Indeed, three-body resonant states are known to exist in many fields of physics, $e.g.$ cold-atom physics~\cite{naidon2017}, nuclear~\cite{nielsen2001a} and hypernuclear physics~\cite{Belyaev2008,Wu2020} and three-body analog systems of excitons in bulk semiconductors~\cite{kazimierczuk2014,belov2022}. For instance, the celebrated Efimov states~\cite{Efimov1970}, first observed in systems of ultracold atoms~\cite{Kraemer2006}, are often in fact resonances and not true bound states~\cite{naidon2017}. How the lifetimes of Efimov three-body resonances are changed, has already been analysed when approaching the three-body dissociation threshold~\cite{nielsen2002}, or in the limit of infinitely large mass ratios~\cite{penkov1999}. Apart from Efimov states, it is relevant to study the stability of three-body resonances, $e.g.$ for the application of condensates of three-body states~\cite{braaten2003,musolino2022}. Moreover, there has been recent progress in the controlled collision of ultracold atoms and molecules~\cite{yang2022,son2022}, and the associated analysis of resonance spectra~\cite{park2023}.

In the present article we theoretically study the lifetimes of three-body resonances against decay into a deep dimer and an unbound particle. In particular, we analyse the dependence on the mass ratio between two different components in our system in the range $1/20 \ldots 20$, which is most relevant for mixtures of ultracold atoms. We employ the Gaussian expansion method (GEM)~\cite{hiyama2003,hiyama2012} together with the complex scaling method (CSM)~\cite{simon1973,moiseyev1998} to obtain the widths of three-body resonances from first principles. Our investigation unveils an intriguing dependence of the resonance widths on the mass ratio. As the mass ratio increases, the widths display an oscillatory behaviour on top of an overall decreasing trend. Notably, particular mass ratios result in a vanishing resonance width, indicating that the resonance becomes in fact stable, which can be interpreted as so-called bound states in the continuum (BIC)~\cite{hsu2016}. We also consider an analytical formula~\cite{penkov1999} for the widths, derived in the limit of large mass ratios, and find that it is unable to reproduce our findings. Despite the fact that the mass ratio is usually not a tunable parameter, it differs depending on the explicit choice of atomic species, nuclei or semiconductor materials.

Moreover, motivated by a previous work~\cite{happ2022}, in which excited states in a one-dimensional (1D) configuration were analysed under the bound-state approximation, we perform calculations for both 1D and 3D, and contrast the results against each other. In particular, for a mass ratio near $22$ (Cs-Li mixture) we find almost vanishing widths of the resonances, justifying the bound-state approximation in the previous work. Comparison of the results for 1D and 3D indicates qualitatively similar behaviour. Our one-dimensional results are particularly interesting for experimental systems with controlled low dimensionality, realised since the advent of optical dipole traps~\cite{bloch2005}.

Our article is organised as follows: In section \ref{sec:systemandmethods}, we introduce the three-body system under consideration together with the employed methods. Then, we present the results of our calculations in section \ref{sec:results}, before discussing our findings and presenting an outlook in section \ref{sec:discussion}.

\section{System and Methods}\label{sec:systemandmethods}
In this section we introduce the three-body system, the regime of interest, and the relevant quantities and methods we use for its analysis.

\begin{figure}[htb]
    \centering
    \begin{subfigure}[c]{0.48\textwidth}
        \centering
        \includegraphics[width=0.75\textwidth]{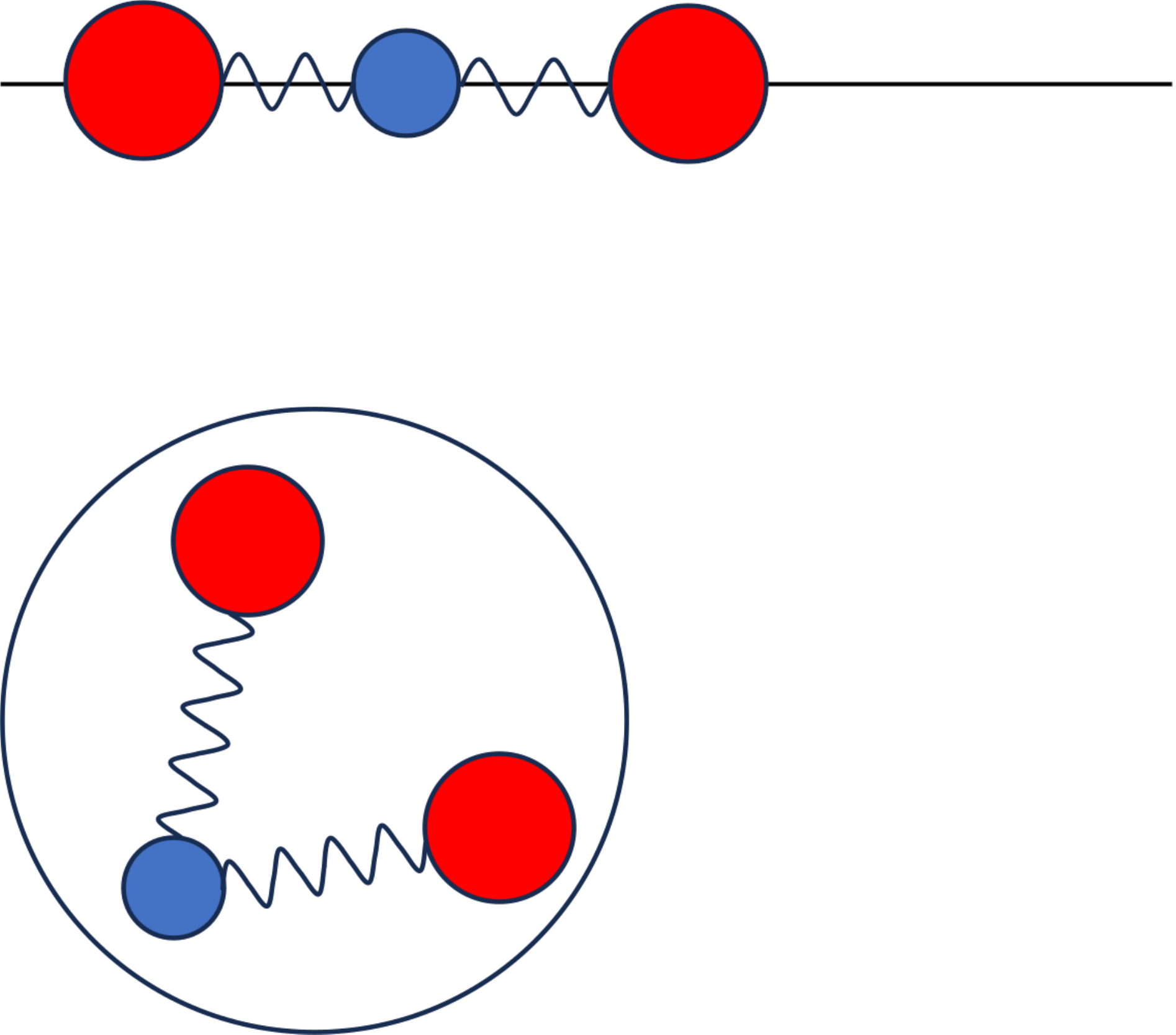}
        \end{subfigure}
    \hfill
    \begin{subfigure}[c]{0.48\textwidth}
        \centering
        \includegraphics[width=\textwidth]{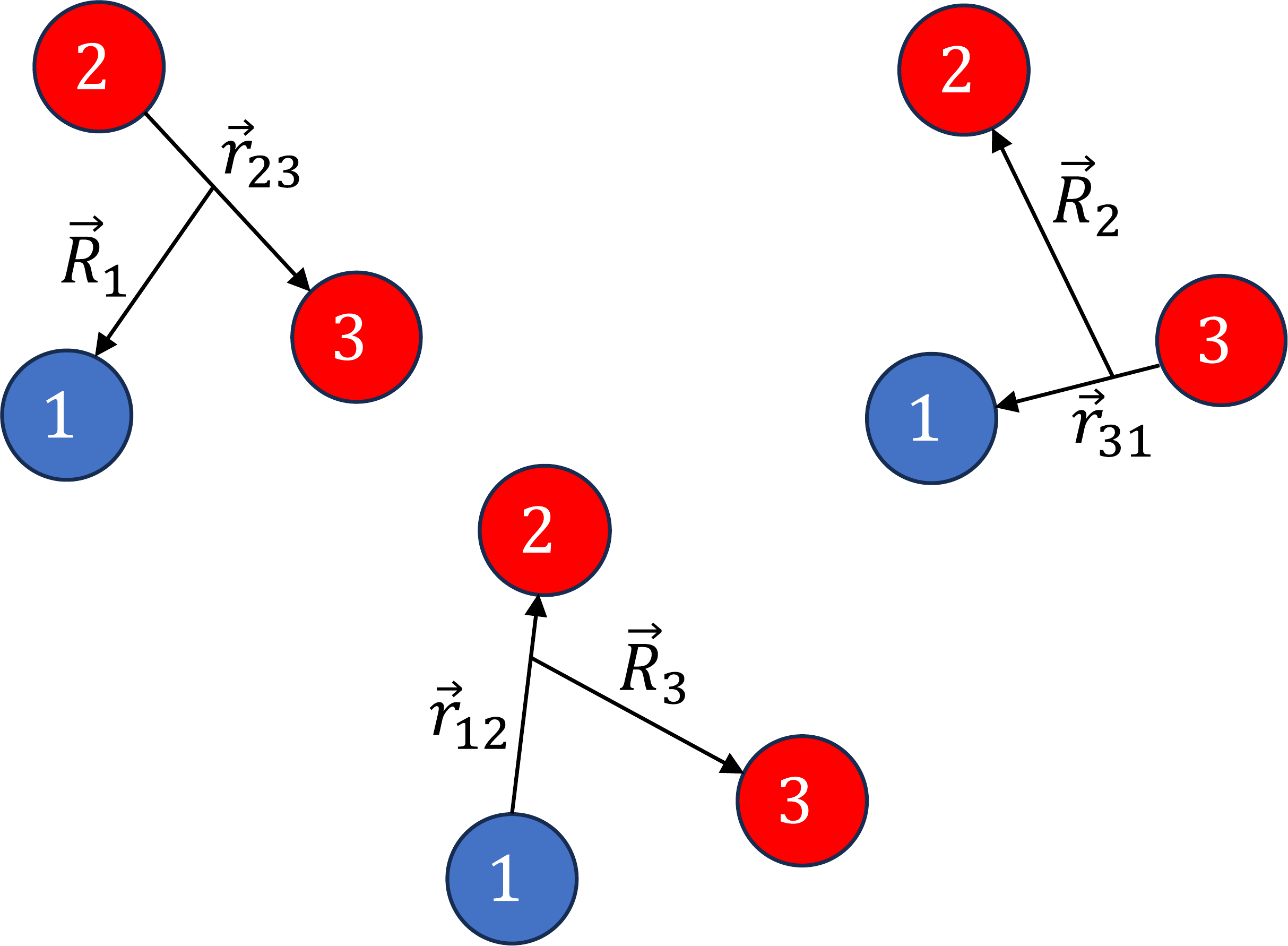}
    \end{subfigure}
    \caption{Left: Configuration of particles and interactions of our three-body system in 1D (top) and 3D (bottom). Interactions are represented by wavy lines. Right: Three sets of Jacobi coordinates. In both subfigures, blue disks indicate the different particle (1) and red disks the two identical bosonic particles (2,3).}
    \label{fig:config_jacobi}
\end{figure}

\paragraph{The three-body system}
Our three-body system consists of two identical bosons (B) and a third, different particle (X). We consider only pair interactions between the different particles, but none between the identical bosons. This is to keep the complexity and the number of system parameters as low as possible, while at the same time ensuring rich enough features.  We consider this three-body system in both a one-dimensional (1D) and three-dimensional (3D) configuration, as depicted schematically in Fig. \ref{fig:config_jacobi} (left). Here, the identical bosons and the third particle are depicted as red and blue disks respectively, and wavy lines indicate the interactions.

\paragraph{Jacobi coordinates}
In order to describe this three-body system we employ the commonly used Jacobi-coordinates~\cite{greene2017}. Since the system is translationally invariant, we can choose to work in the centre-of-mass frame of the three-body system, and conveniently set the centre-of-mass coordinate $\vec{\mathcal{R}}=0$. The remaining two relative coordinates are defined by 
\begin{align}
    \vec{r}_{ij} \equiv \vec{r}_j - \vec{r}_i,\qquad \qquad \vec{R}_{k} \equiv \vec{r}_k - \frac{m_i \vec{r}_i + m_j\vec{r}_j}{m_i + m_j}
\end{align}
and depicted in Fig. \ref{fig:config_jacobi} (right). Here, $\vec{r}_{i}$ and $m_i$ respectively denote the absolute coordinate and the mass of particle $i$. In our case we denote the distinct particle by the number 1, and the two identical bosons by 2 and 3. In total there are three sets of Jacobi coordinates, hence we are free to choose $\{i,j,k\} = \{1,2,3\}$, or cyclic permutations thereof. The definition of Jacobi coordinates is identical in 1D and 3D, however in the former case, the vectorial character of the coordinates reduces to simple scalars. In order to keep the notation simple, we also denote the 1D coordinates by vectors.

\paragraph{Model}
The centre-of-mass Schr\"odinger equation governing this three-body system reads
\begin{equation}
    \left[-\frac{\hbar^2}{2\mu_{ij}}\nabla_{\vec{r}_{ij}}^2 -\frac{\hbar^2}{2\mu_{k}}\nabla_{\vec{R}_{k}}^2 + V(r_{12}) + V(r_{31})\right] \Psi(\vec{r}_{ij},\vec{R}_k) = E \Psi(\vec{r}_{ij},\vec{R}_k),
\end{equation}
where we introduced the reduced masses
\begin{align}
\mu_{ij} \equiv \frac{m_i m_j}{m_i + m_j},\qquad \qquad \mu_{k} \equiv \frac{m_k (m_i + m_j)}{m_k + m_i + m_j},
\end{align}
and the distances $r_{ij} \equiv |\vec{r}_{ij}|$. Again, in the 1D case the Laplacians reduce to scalar second derivatives.

We consider purely attractive model interactions of Gaussian shape
\begin{equation}
    V(r) = v_0 e^{-(r/r_0)^2},
\end{equation}
which can be used to characterise different physical systems~\cite{alvarez-rodriguez2016,deltuva2020} by tuning of the two parameters $v_0<0$ and $r_0$. Since we consider isotropic interactions, they depend only on the distance $r = |\vec{r}|$ between the interacting particles.

For a uniform description of physical systems that may live in vastly different energy ranges, it is useful to introduce dimensionless variables which describe the lengths, masses, and energies in units of appropriately chosen characteristic quantities. Since our three-body system is determined by the two-body interactions, we scale the masses 
\begin{equation}\label{eq:primescaling_mass}
    m_i' \equiv \frac{m_i}{2\mu_{bx}}
\end{equation}
by the reduced mass $\mu_{bx}$ ($x=1$, $b=2,3$) of the two interacting particles, and the lengths 
\begin{equation}\label{eq:primescaling_length}
    \begin{pmatrix}
        \vec{r}_{ij}' \\
        \vec{R}_k'
    \end{pmatrix} \equiv \frac{1}{r_0} \begin{pmatrix}
        \vec{r}_{ij} \\
        \vec{R}_k
    \end{pmatrix}
\end{equation}
by the range $r_0$ of interaction. Consequently, the energy
\begin{equation}\label{eq:primescaling_energy}
    E' \equiv \frac{1}{E_\mathrm{char}} E,
\end{equation}
is rescaled in units of the characteristic energy $E_\mathrm{char} \equiv \hbar^2/(2\mu_{bx}r_0^2)$. Moreover, the reduced masses then become functions of a single parameter only, which is the mass ratio $\beta \equiv m_b/m_x = m_2 / m_1 = m_3/m_1$ between the two components.

Accordingly, using these rescaled quantities, the two-body system of interacting particles $x=1$ and $b=2$ (or $b=3$) is governed by the Schr\"odinger equation
\begin{equation}\label{eq:2bsgl}
    \left[-\nabla_{\vec{r}_{bx}'}^2 + V'(r'_{bx}) \right] \psi(\vec{r}_{bx}') = E^{(2)\prime} \psi(\vec{r}_{bx}'),
\end{equation}
with
\begin{equation}
    V'(r') = \frac{v_0}{E_\mathrm{char}} e^{-r'^2} \equiv v_0' e^{-r'^2}.
\end{equation}
We note that within this notation the relation between $v_0'$ and $E^{(2)\prime}$ becomes independent of the masses of the constituent particles, and therefore also independent of the mass ratio.

\begin{figure}[htb]
    \centering
    \begin{subfigure}[c]{0.48\textwidth}
        \centering
        \includegraphics[width=\textwidth]{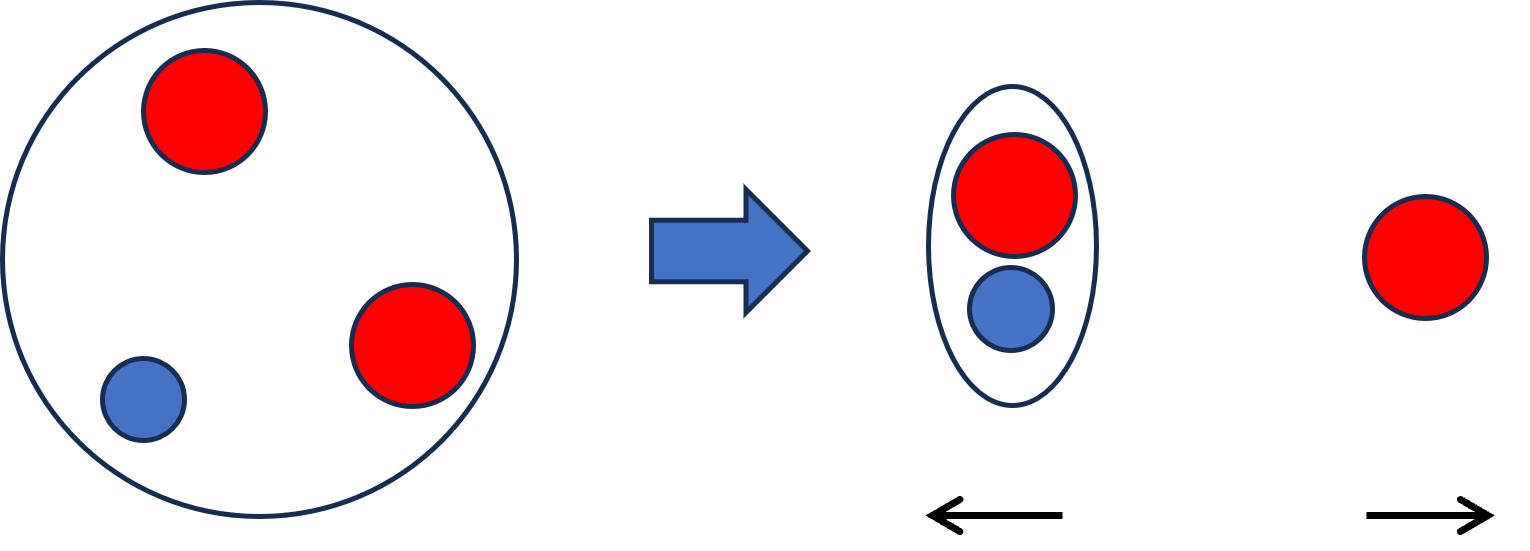}
    \end{subfigure}
    \hfill
    \begin{subfigure}[c]{0.48\textwidth}
        \centering
        \includegraphics[width=\textwidth]{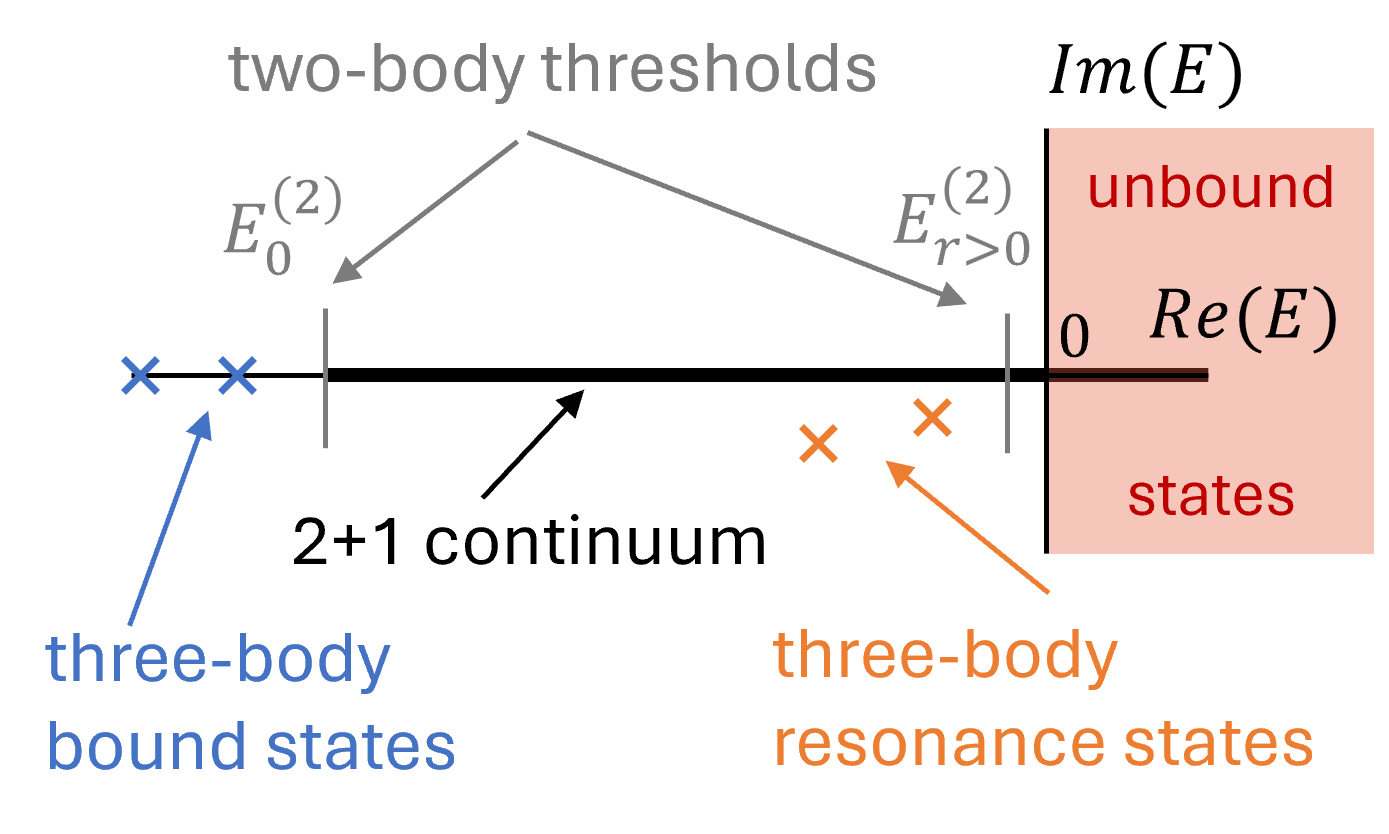}
    \end{subfigure}
    \caption{Left: Schematic representation of the decay process of three-body resonances into the continuum of a deeply bound two-body state and an unbound particle. Due to energy and momentum conservation the two resulting parts fly away in opposite directions, as indicated by the arrows. Right: Schematic energy spectrum of our three-body system. The two-body ground state energy $E_0^{(2)}$ (grey) divides the three-body spectrum into bound states (blue) which lie below it, and resonant states (orange) which lie above it. Above the dissociation threshold $E=0$, there can be unbound states (red shaded area). Moreover, continua (black thick line) of a dimer state together with an unbound particle start above the binding energy of each two-body bound state (grey).}
    \label{fig:schematic}
\end{figure}

\paragraph{Regime of interest}
We are interested in three-body resonances which are characterised by the possibility to decay into continuum states of a deeply bound dimer together with an unbound particle, see Fig. \ref{fig:schematic} (left). Depending on the specific physics community, this process is sometimes referred to as \textit{predissociation}, \textit{Auger effect}, etc.~\cite{Newton1982,l.d.landau1958}.

In Fig. \ref{fig:schematic} (right) we schematically present a typical three-body spectrum. It consists of true three-body bound states (blue) below the deepest two-body energy level $E_0^{(2)}$ (grey), and three-body resonances (orange) above it. While the bound states are stable, as there is no continuum state to decay into, the resonances are only metastable and sometimes referred to as quasi-bound states. Each two-body binding energy marks the start of a continuum of states (thick black line) consisting of a bound pair with that energy, together with an unbound third particle. Above the dissociation threshold (red shaded area), there can be unbound states of three free particles. 

The resonances are characterised by a complex-valued eigenenergy~\cite{Kukulin1989}
\begin{equation}
    E = E_{\mathrm{r}} - \frac{i}{2}\Gamma
\end{equation}
with resonance position $E_{\mathrm{r}}$, and width $\Gamma$. The width can be related~\cite{moiseyev1998} to the time-dependent probability density
\begin{equation}
    \left|\Psi_{\mathrm{res}}(r,t)\right|^2 = \left|\Psi_{\mathrm{res}}(r)\right|^2 e^{-\Gamma t /\hbar} \equiv \left|\Psi_{\mathrm{res}}(r)\right|^2 e^{-t / \tau},
\end{equation}
and then the imaginary part of the energy to the lifetime
\begin{equation}\label{eq:gammatau}
    \tau \equiv \frac{\hbar}{\Gamma} = -\frac{\hbar}{2\,\mathrm{Im}(E)}.
\end{equation}

For a uniform description of the three-body system for all mass ratios, we fix the rescaled energy $E^{(2)\prime}$ of the first excited $s$-wave state to the value $E^{(2)\prime} = -0.1$. In 1D we fix the energy of the first excited symmetric state, respectively. This corresponds to a scattering length of about three times the interaction length, and is a regime which is feasible in both nuclear and cold atom physics. We choose to study the regime of the positive side of the scattering length (the corresponding two-body state is bound) in order to have three-body states in both 3D and 1D. While in 3D the three-body states exist also on the negative side of the scattering length, such so-called Borromean states are not known to exist in 1D, at least not for purely attractive pair-interactions~\cite{schnurrenberger2024}.

Above in Eq. \eqref{eq:2bsgl}, we have demonstrated that the rescaled two-body system becomes independent of the masses. Hence, we can obtain the desired value of $E^{(2)\prime}$ by choosing $v_{0}' = -19.77$ in 3D and $v_{0}' = -5.44$ in 1D for all values of the mass ratio. As a result, the two-body spectrum remains constant under change of the mass ratio and hence any residual effect is a pure three-body effect.

\paragraph{Method}
For solving both the two-body and the three-body problem, we employ the Gaussian expansion method (GEM)~\cite{hiyama2003,hiyama2012}. We expand the total three-body state as
\begin{equation}
    |\Psi\rangle = \sum_{c=1}^{3}\sum_{\alpha=1}^{\alpha_\mathrm{max}} A_\alpha^{(c)} |\Phi_\alpha^{(c)}\rangle,
\end{equation}
where $c$ and $\alpha$ respectively run over the different Jacobi sets, and the basis functions $|\Phi_\alpha^{(c)}\rangle$. $\vec{A}^{(c)}$ denotes the vector of coefficients. In 3D we apply the method exactly as described in the review article~\cite{hiyama2003} with
\begin{equation}
    \langle\vec{r}_c,\vec{R}_c|\Phi_{\alpha}^{(c)}\rangle = \Phi_{\alpha}^{(c)}(\vec{r}_c,\vec{R}_c) = \phi_{n_{\alpha},l_{\alpha}}^{(c)}(r_c) \psi_{N_{\alpha},L_{\alpha}}^{(c)}(R_c) Y_{l_{\alpha},m_{\alpha}}(\hat{r}_c) Y_{L_{\alpha},M_{\alpha}}(\hat{R}_c),
\end{equation}
where $Y_{l,m}$ are the spherical harmonics, and $\{l_{\alpha},m_{\alpha}\}$ and \{$L_{\alpha},M_{\alpha}\}$ denote the angular momentum and its projection for the relative coordinates $\vec{r}$ and $\vec{R}$ respectively. Moreover, we have employed the simplified notation, where now $e.g.$ $\vec{r}_{c=1}$ is the Jacobi coordinate $\vec{r}_{23}$, and $\hat{r}$ denotes the directional unit vector $\vec{r}/r$. For the 1D case we establish the GEM in a similar way with
\begin{equation}
    \langle z_c,Z_c|\Phi_{\alpha}^{(c)}\rangle = \Phi_{\alpha}^{(c)}(z_c,Z_c) = \phi_{n_{\alpha},l_{\alpha}}^{(c)}(z_c) \psi_{N_{\alpha},L_{\alpha}}^{(c)}(Z_c),
\end{equation}
where we drop the angular part, and for clarity use the variables $z_c$ and $Z_c$ to indicate the 1D character.

The GEM is a variational method and relies on expanding the spatial part of the unknown state into a set of basis functions of Gaussian shape
\begin{align}
    \phi_{n,l}(r) =&~ N_{n,l} r^{l} e^{-\nu_n r^2} &~ 1\leq n \leq n_{\mathrm{max}},&~\qquad 1\leq l \leq l_{\mathrm{max}} \\
    \Phi_{N,L}(R) =&~ N_{N,L} R^{L} e^{-\lambda_N R^2} &~ 1\leq N \leq N_{\mathrm{max}},&~\qquad 1\leq L \leq L_{\mathrm{max}}
\end{align}
with normalisation factors $N_{n,l}$ and $N_{N,L}$. Due to the scalar arguments we can use the same form of basis functions also for the 1D case. Overall, due to the non-orthogonal character of the Gaussian basis functions, this creates a generalised matrix eigenvalue problem and both the eigenenergies and corresponding eigenvectors $\vec{A} = (\vec{A}^{(1)},\vec{A}^{(2)},\vec{A}^{(3)})$ are then obtained by diagonalisation. Since our system contains two identical particles, and no interaction between them, we can reduce our analysis to a single set of Jacobi coordinates ($c = 2$ or $c = 3$).

In 3D, it turns out that the lowest three-body resonances corresponding to the $2s$ two-body bound states lie already far below ($E' \simeq -3$) the $1p$ two-body energy ($E^{(2)\prime} \simeq -0.25$) owing to their (strong) borromean character. By this argument we justify neglecting the influence of this two-body threshold lying far above the three body energies and hence choose $l_{\mathrm{max}} = L_{\mathrm{max}} = 0$. On the contrary, in 1D the three-body resonances corresponding to excited symmetric two-body states lie above ($E' \simeq -0.2$) the odd-wave two-body energy ($E^{(2)\prime} \simeq -1.5$). Hence, we cannot neglect the corresponding continuum and have to choose $l_{\mathrm{max}} = L_{\mathrm{max}} = 1$ in 1D.

Moreover, for the 3D case, we use the complex-ranged Gaussian basis functions~\cite{hiyama2003}, for which the Gaussian ranges are transformed to $\nu_n \to (1\pm i\omega)\nu_n$, and accordingly for $\lambda_N$. As a result, the number of basis functions doubles. A value of $\omega=0.8$ has yielded good results. The other numerical parameters are listed in Table \ref{tab:numparams}.

\begin{table}[htb]
    \centering
    \begin{tabular}{c|c|c|c|c|c|c|c}
         & $n_\mathrm{max} = N_\mathrm{max}$ & $l_\mathrm{max} = L_\mathrm{max}$ & $\alpha_\mathrm{max}$ & $\nu_1$ & $\nu_\mathrm{max}$ & $\lambda_1$ & $\lambda_\mathrm{max}$\\
    \midrule
    1D & 32 & 1 & 2048 & 318.9 & 0.037 & 45.65 & 0.023\\
    3D & $2 \times 16$ & 0 & 1024 & 68.83 & 0.0058 & 61.85 & 0.011
    \end{tabular}    
    \caption{Numerical parameters of the GEM used in our calculations. The total number of basis functions $\alpha_\mathrm{max}=\left(n_\mathrm{max}(l_\mathrm{max}+1)\right)^2$ is determined by $n_\mathrm{max}$ and $l_\mathrm{max}$. In the 1D case, $\alpha_\mathrm{max}$ can be reduced from 4096 to 2048 due to global parity conservation. Depending on the mass ratio, these number of basis functions, as well as minimal ($\nu_1$, $\lambda_1$) and maximal ($\nu_\mathrm{max}$, $\lambda_\mathrm{max}$) Gaussian ranges yield an accuracy of the resonance widths between $10^{-3}$ and $10^{-6}$.}
    \label{tab:numparams}
\end{table}

We make use of the complex scaling method (CSM)~\cite{moiseyev1998} to obtain the eigenenergies of the three-body resonances. This method is based on introducing a scaling
\begin{equation}
    \begin{pmatrix}
        \vec{r} \\
        \vec{R}
    \end{pmatrix}
    \to e^{i\theta}
        \begin{pmatrix}
        \vec{r} \\
        \vec{R}
    \end{pmatrix}
\end{equation}
to the coordinates. By solving the Schr\"odinger equation using this complex scaling, we can obtain the complex eigenenergies by ordinary numerical methods for bound states.
Within the CSM, the discrete spectrum of bound and resonant states remains unchanged, while the various continua are rotated by an angle $-2\theta$ in the complex space of energies, see Fig. \ref{fig:CSMPlots}. When the rotation angle is chosen large enough this uncovers the resonances. For more details we refer to Refs.~\cite{simon1973,moiseyev1998,Lazauskas2023}.

\section{Results and Discussion}\label{sec:results}

\subsection{Complex-rotated energy spectra}
\begin{figure}[htb]
    \centering
    \includegraphics[width=\textwidth]{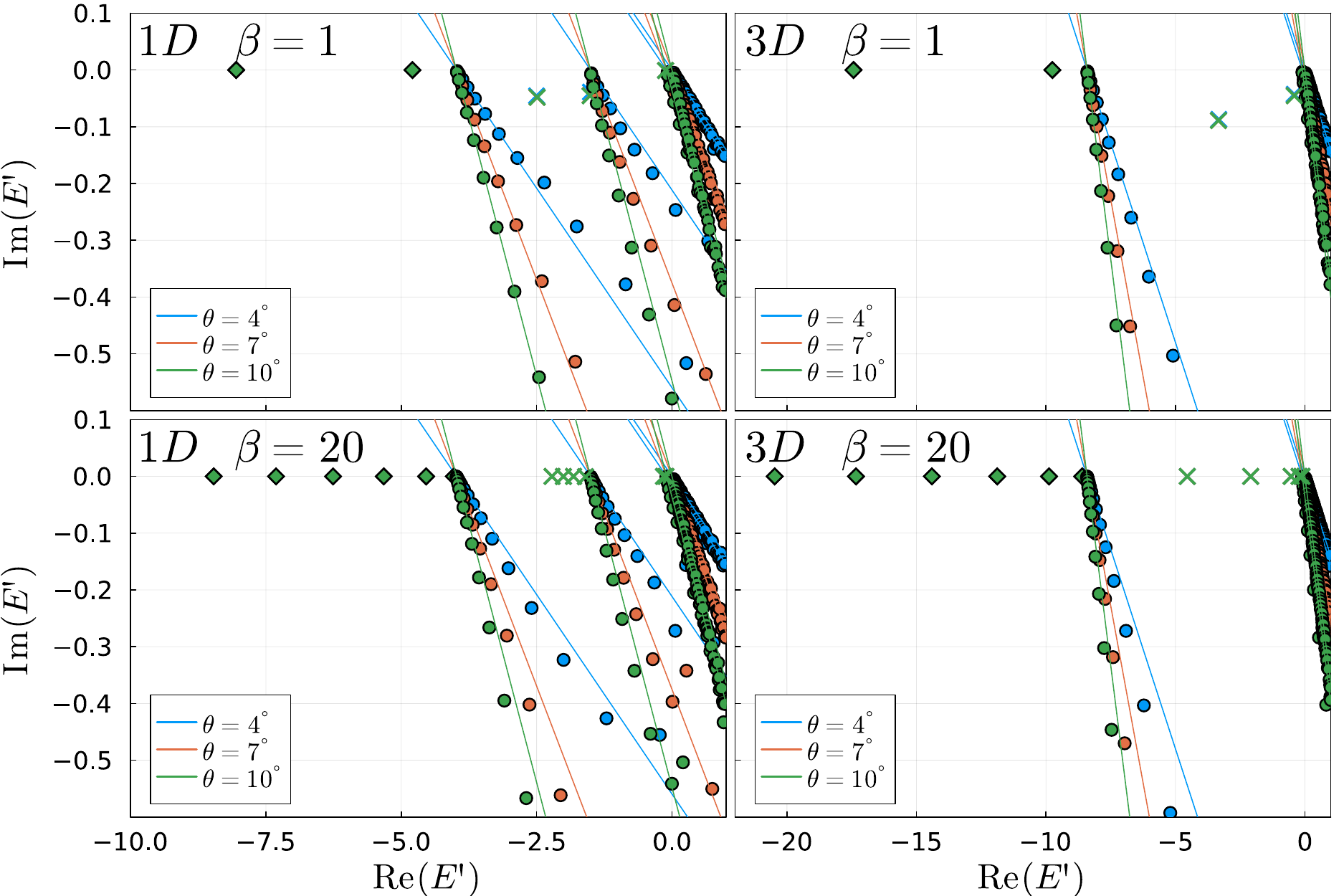}
    \caption{Plane of complex energies with the complex-rotated three-body energy spectrum for 1D (left column) and 3D (right column), divided also in $\beta=1$ (top row) and $\beta=20$ (bottom row). The calculations are performed for three different complex-rotation angles $\theta=4^{\circ}$ (blue), $7^{\circ}$ (orange), and $10^{\circ}$ (green). The discretised continuum states (filled circles) of a deep dimer together with an unbound particle start above each two-body threshold and appear along a line rotated by $2\theta$ downwards. In contrast, the discrete spectrum of bound (diamonds) and resonant states (crosses) remains constant for the different angles. For the mass ratio $\beta=1$ (top row), in both 1D and 3D the three-body resonances have a sizeable imaginary part. Increasing the mass ratio to $\beta=20$ (botom row) reduces the imaginary part significantly to basically zero, hence the resonances become almost indistinguishable from bound states. Again, this can be seen for both 1D and 3D.}
    \label{fig:CSMPlots}
\end{figure}
We analyse the resonances in the range of mass ratios $1/20 \leq \beta \leq 20/1$. This range covers the most typical mass ratios in systems of ultracold atoms for both two light bosons and one heavy particle ($\beta <1$), as well as for the reverted case of two heavy bosons and a light particle ($\beta >1$). Moreover, our results indicate that this range is where most of the observed effect is taking place.

The resulting three-body spectra are presented in Fig. \ref{fig:CSMPlots}. The four diagrams are divided into two columns, where the left (right) column shows the result for 1D (3D), and two rows where the top (bottom) row depicts the case of mass ratio $\beta=1$ ($\beta = 20$). In each diagram the horizontal and vertical axes respectively represent the real and imaginary part of the three-body energy. The three colours indicate calculations for different rotation angles $\theta=4^{\circ},\,7^{\circ},\,10^{\circ}$. We have added solid lines as a guide to the eye for the corresponding rotation angles $-2\theta$ in energy space.

As previously explained, within the CSM the discretized-continuum states (filled circles) appear at a rotated position. We note the existence of several continua starting at the different bound states of the BX subsystem, as well as at the dissociation threshold $E=0$. Again, we emphasise that within the rescaled variables, the two-body energies are independent of the mass ratio, hence they are at the same position in both rows of diagrams. In contrast to the rotated continua, the bound (diamonds) and resonant (crosses) states remain unchanged under the rotation. Overall, the results show good convergence under change of the complex-rotation angle $\theta$.

We see that in the top row of diagrams ($\beta=1$) the resonances are located visibly away from the real energy axis, while in the bottom row ($\beta=20$) they all practically lie on it. This indicates that for increased mass ratio $\beta$, the imaginary parts of the resonant states' eigenenergies decrease and they become almost indistinguishable from bound states. This is in agreement with a previous conjecture~\cite{happ2022} made for the 1D case and $\beta=20$ (bottom left diagram), based solely on a real-valued analysis of the resonance position $E_\mathrm{r}$ (bound-state approximation). Overall, we find qualitatively the same behaviour for both the 1D and 3D case. Moreover, we note that with increased mass ratio both the number of three-body bound states and resonances increase. For bound states in 1D this dependence has already been studied in more detail in Refs.~\cite{Kartavtsev2009,happ2019a}.

\subsection{Energy-width of three-body resonances}
\begin{figure}[h!]
    \centering
    \begin{subfigure}[b]{0.7\textwidth}
        \centering
        \includegraphics[width=\textwidth]{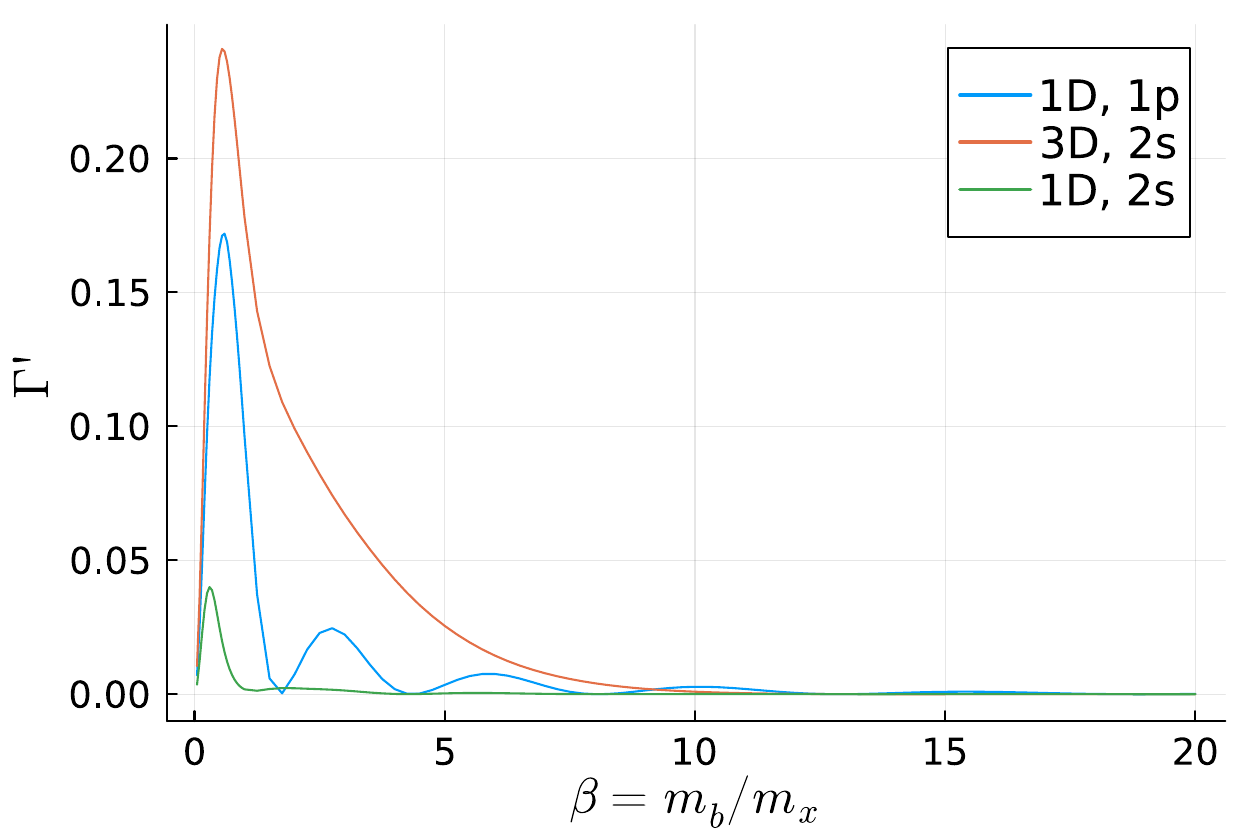}
    \end{subfigure}
    \\
    \begin{subfigure}[b]{0.7\textwidth}
        \centering
        \includegraphics[width=\textwidth]{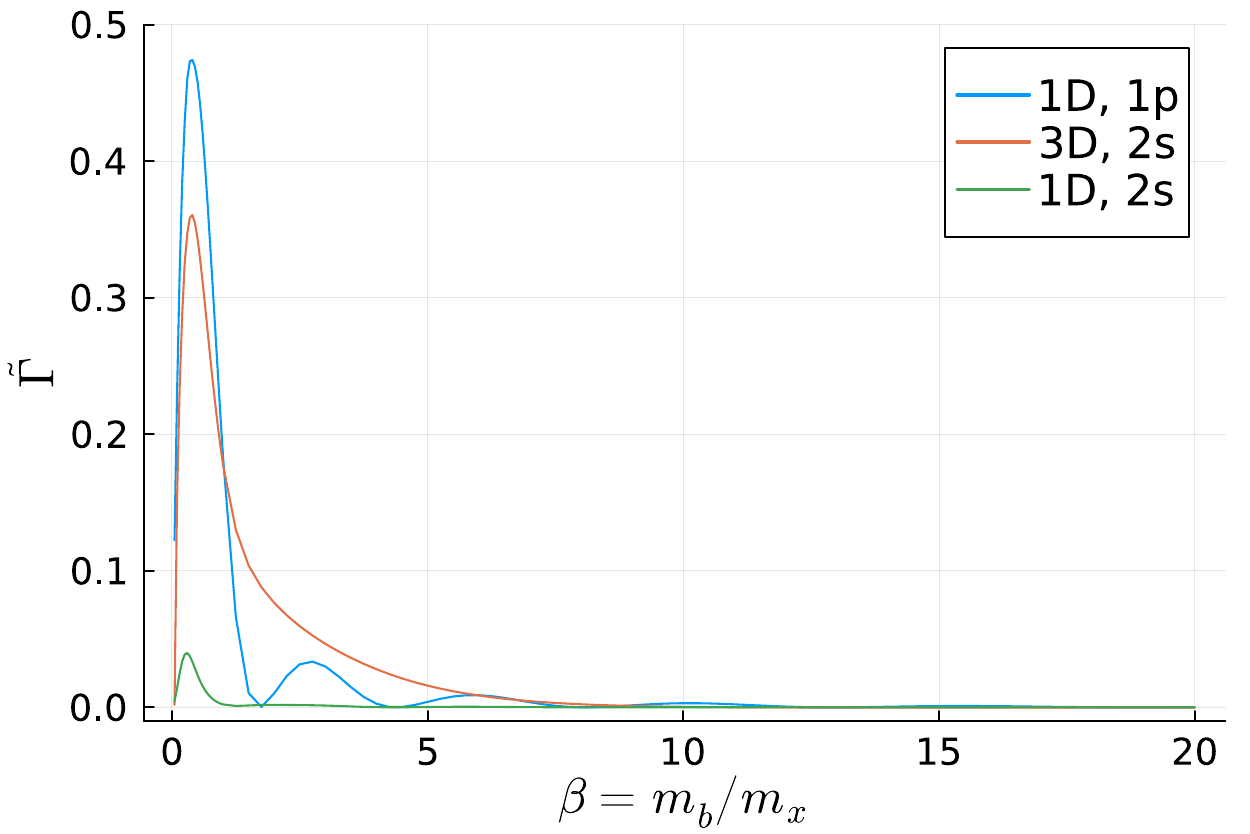}
    \end{subfigure}
    \caption{Top: Widths $\Gamma'$ of three-body resonances as a function of the mass ratio $\beta$. Here we depict the deepest one of each ``family'' of resonances associated with a particular BX state: $(1D, 1p)$ (blue), $(1D, 2s)$ (green), $(3D, 2s)$ (orange). For all states we find that for increasing mass ratios, $\Gamma'$ oscillates on top of an overall decrease. However, the frequency and amplitude differs from state to state. It can be seen most clearly for the $(1D,1p)$ resonance which shows several maxima and minima in the considered regime. In comparison, $\Gamma'$ for the $(1D,2s)$ resonance is overall smaller, such that the oscillations are hard to see. For the $(3D,2s)$ resonance there is only a single maximum visible in the region of interest (around the equal mass case), however a minimum at $\beta \simeq 15$ can be inferred from Fig.~\ref{fig:TauvsMR}. Bottom: the same result for $\tilde{\Gamma}$, based on another scaling, Eq. \eqref{eq:tildescaling}, of energies with the different particle's mass $m_x$. The effect remains, however the amplitudes of oscillations are slightly changed.}
    \label{fig:ImEprimeandtilde}
\end{figure}
In Fig. \ref{fig:ImEprimeandtilde} (top) we quantitatively analyse the resonance widths $\Gamma'$ as a function of the mass ratio. Here we focus on three resonant states, which are each the deepest, i.e. the ones with the lowest $E_r$, of each ``family'' of resonances associated with a particular state in the BX subsystem.  We refrain from analysing the higher excited resonances in each family as they partially exist only within a small range of mass ratios. The three-body resonance labeled ($1D, 1p$) lies between the ground and first excited state of the BX system, whereas the $(1D, 2s)$ resonance lies above the first excited and below the second excited BX state (see Fig. \ref{fig:CSMPlots}). Even though in 1D no partial waves exist and the two-body states instead have well defined parity ($e.g.$ even and odd), we keep the notation of $s$ (even) and $p$ (odd) in similarity to the 3D case. There, we analyse only the $(3D, 2s)$ state as explained above.

We have already seen in the previous subsection that for mass ratios larger than one, the resonance width $\Gamma'$ decreases. The more detailed analysis in Fig.~\ref{fig:ImEprimeandtilde} however reveals an even richer dependence: we find that with increasing mass ratios, $\Gamma'$ displays a damped-oscillatory behaviour, with a global maximum near the equal mass case. The frequency and amplitude of oscillations depends however on the particular resonance. Whereas they can be clearly seen for the $(1D,1p)$ resonance, the amplitude is more suppressed for the $(1D,2s)$ resonance. For the $(3D,2s)$ resonance the oscillation-frequency is smaller such that we see only a single maximum in the region of analysed mass ratios.

The oscillations come with the remarkable feature that the resonance width vanishes entirely for specific values of the mass ratio. This indicates the possibility for so-called bound states in the continuum (BIC)~\cite{hsu2016} to be present in few-body systems. Even though in most experiments of cold atoms or nuclei the mass ratio is not a directly tunable parameter, it can be varied e.g. for excitons in semiconductor systems \cite{kazimierczuk2014}.

In order to rule out that our effect is solely an artifact of the specific global scaling, Eqs. \eqref{eq:primescaling_mass}-\eqref{eq:primescaling_energy}, we introduce another scaling
\begin{align}\label{eq:tildescaling}
    \tilde{E} \equiv \frac{m_x r_0^2}{\hbar^2} E = 2\frac{m_{x}}{\mu_{bx}}E'.
\end{align}
Here, instead of scaling with the reduced mass $\mu_{bx}$ of interacting particles, we scale with $m_{x}$, the mass of the non-identical particle. Moreover, we now fix $\tilde{E}^{(2)} = -0.1$ instead of $E^{(2)\prime}$. The corresponding result is presented in Fig. \ref{fig:ImEprimeandtilde} (bottom). The main difference to Fig. \ref{fig:ImEprimeandtilde} (top) is that the amplitude of oscillations changes a bit depending on the particular resonant state. Crucially, the main effect persists in its essential form.

\begin{figure}[htb]
        \centering
        \includegraphics[width=0.7\textwidth]{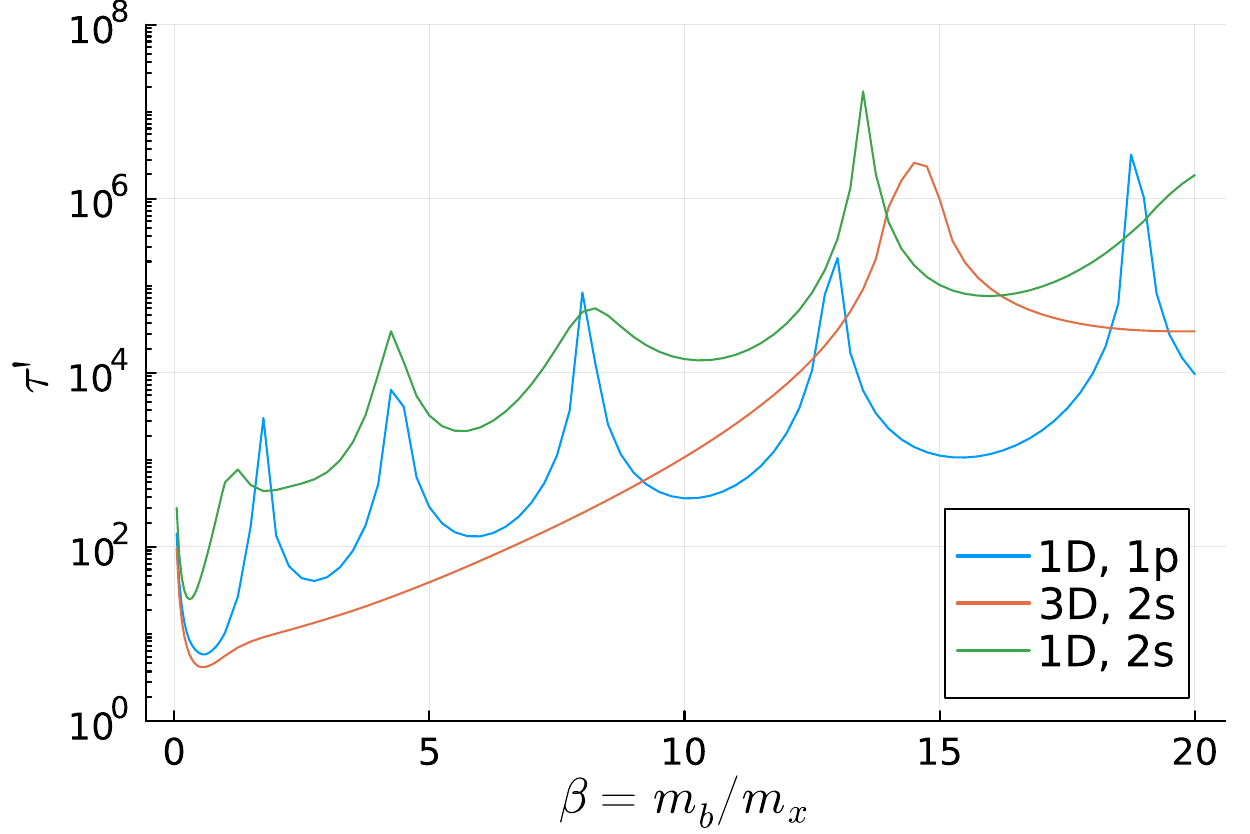}
        \caption{Lifetime $\tau'$, Eq. \eqref{eq:tauprime}, in log-scale as a function of the mass ratio $\beta$. Due to the reciprocal relation, Eq. \eqref{eq:gammatau}, the maxima in Fig. \ref{fig:ImEprimeandtilde} show here as minima, and vice-versa. For all the considered states, we find that the lifetimes increase by several orders of magnitude between $\beta=1$ and $\beta =20$, indicating strongly increased stability of the resonances. On top of that we find several special mass ratios for which the resonance width vanishes, i.e. the lifetime diverges. The maximum of the peaks in $\tau'$ is limited by our numerical accuracy which varies between $10^{-3}$ ($\beta \simeq 1$) and $10^{-6}$ ($\beta \simeq 20$), depending on the mass ratio.}
        \label{fig:TauvsMR}
\end{figure}

For a better representation of the different orders of magnitude in play, we display in Fig. \ref{fig:TauvsMR} the lifetime
\begin{equation} \label{eq:tauprime}
    \tau' \equiv \frac{\tau}{\tau_\mathrm{char}}
\end{equation}
in log-scale against the mass ratio. Hence, the lifetimes are shown in units of a characteristic lifetime $\tau_\mathrm{char} \equiv \mu_{bx}r_0^2/\hbar$. For $e.g.$ a system of ultracold atoms with an estimated interaction range $r_0 \simeq 100 a_0$, i.e. 100 Bohr radii, $\tau_\mathrm{char}$ is of the order of nanoseconds (ns). Indeed, a recent experiment colliding ultracold atoms with molecules has reported a lifetime of around 60\,ns~\cite{son2022}.

Due to the relation \eqref{eq:gammatau} between $\tau$ and $\Gamma$, smaller (larger) resonance widths directly translate into longer (shorter) lifetimes. Here, we see much more clearly the overall trend of increased stability for larger mass ratios together with the oscillatory behaviour. In particular, we see that the frequency of oscillations is quite different between the $1D$ and $3D$ cases. Moreover, the minimum of $\Gamma'$ for the $(3D,2s)$ resonance near $\beta \simeq 15$, not visible in Fig.~\ref{fig:ImEprimeandtilde}, now clearly shows as a maximum. Overall, the height of the peaks is limited by both (i) how densely we sample the mass ratio near the minimum of $\Gamma'$, and (ii) our numerical accuracy, which depends on the mass ratio and lies between around $10^{-3}$ and $10^{-6}$. We estimated the order of magnitude of the numerical accuracy from performing the calculations for several different complex rotation angles $\theta$ and sizes $\alpha_\mathrm{max}$ of the basis set.

\subsection{Comparison to analytical formula}
Already more than two decades ago, Pen'kov has derived~\cite{penkov1999} an analytical formula to approximately describe the dependence of three-body resonance widths as a function of the mass ratio. The formula was derived for the Efimov scenario with three separable, resonant pair interactions. Despite assumed to be resonant, one of the interactions was considered to provide a sufficiently deep two-body state compared to the weakly-bound three-body resonances of interest. Moreover, the formula was derived in the limit of very large mass ratios $m_2/m_1 \to \infty$. Overall, the considered regime and states are therefore quite different from the present work. Nevertheless it is interesting to compare the formula to the results presented here and to check its validity.

\begin{figure}[htb]
        \centering
        \includegraphics[width=0.7\textwidth]{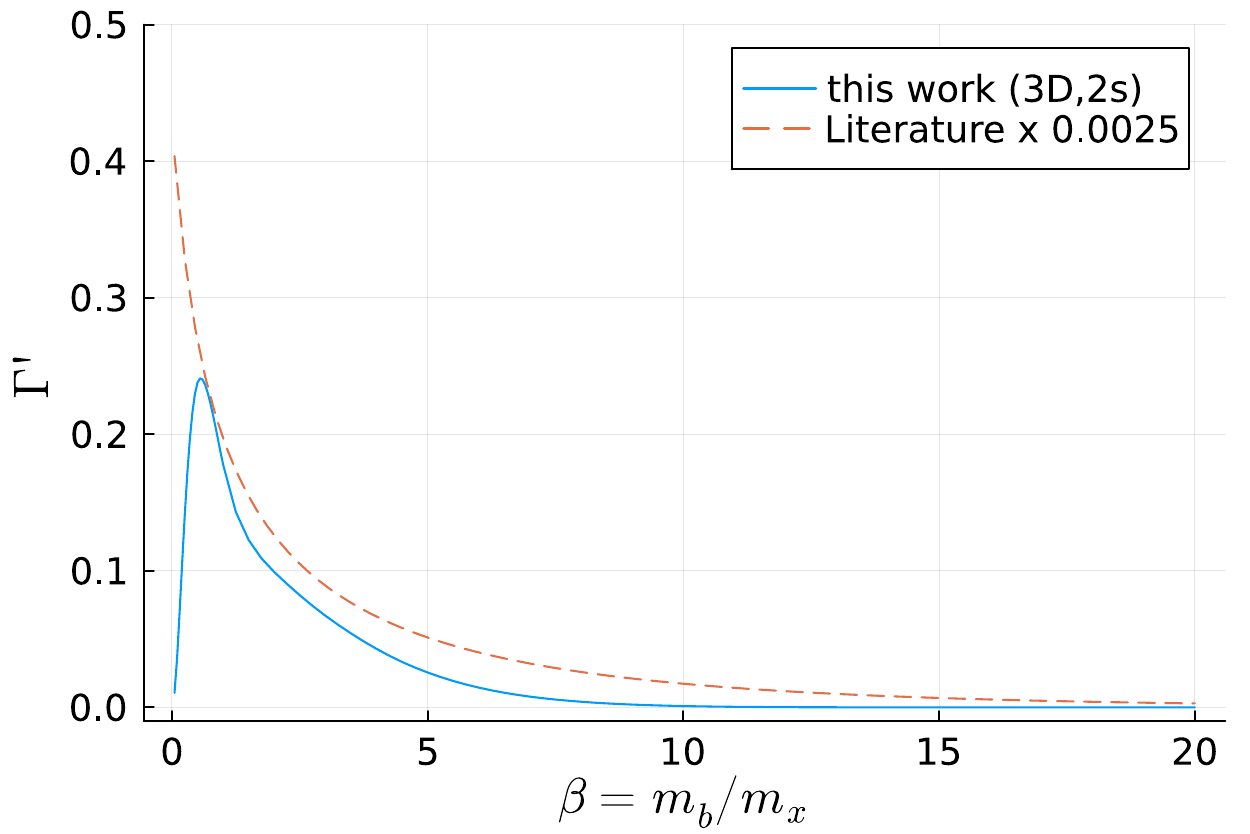}
        \caption{Comparison of our results for the resonance width $\Gamma'$ as a function of the mass ratio $\beta$ with those based on Equation (22) of Ref.~\cite{penkov1999}. The blue line denotes our results for the $(3D, 2s)$ resonance as shown in Figs. \ref{fig:ImEprimeandtilde}, and the orange line the result from the literature multiplied by a factor of $0.0025$. The latter overestimates the resonance width by more than two orders of magnitude, and is unable to reproduce the drop-off for $\beta \lesssim 1$.}
        \label{fig:Penkov}
\end{figure}

In Fig. \ref{fig:Penkov} we display our result for the $(3D, 2s)$ resonance (blue solid line) together with the result using Eq.~(22) of Ref.~\cite{penkov1999} (orange dashed line) in a diagram of $\Gamma'$ against the mass ratio. We highlight here that we have scaled that formula by a factor of $1/400 = 0.0025$ for an easier comparison. Taking this factor into consideration, we see that in the considered range of mass ratios $1/20 \leq \beta \leq 20$, the formula from Ref.~\cite{penkov1999} overestimates the resonance width by more than two orders of magnitude. Moreover, the reduction of $\Gamma'$ for $\beta \lesssim 1$ is missing. We conclude that the formula is not able to describe our findings, which is not surprising considering the fact that it was derived for a specific scenario which is quite different from the one studied here.

Nevertheless, we want to highlight that the original formula of Ref.~\cite{penkov1999} already predicts a damped-oscillatory behaviour of resonance lifetimes in the limit of very large mass ratios. As previously mentioned, the derivation however heavily relied on the presence of the Efimov effect. In our work it is possible that the result for the 3D resonance contains some remainder of the Efimov effect, despite being relatively far away from its typical regime, which might explain the obtained oscillations. On the other hand, we find the damped-oscillatory behaviour also in the 1D case where the Efimov effect is absent.

\section{Summary and Outlook}\label{sec:discussion}
In this article we have shown that the lifetime of three-body resonances strongly depends on the mass ratio between the two components. A larger mass discrepancy in favour of either species results in a stability increase of several orders of magnitude against decay into a continuum of a deeply-bound dimer scattering with an unbound particle. Additionally, the resonance width displays an oscillatory behaviour with the mass ratio, including particular points where the states become exceptionally stable. Employing a scaling in which the two-body spectrum remains constant under change of the mass ratio has allowed us to rule out the effect of shifted two-body thresholds on the stability, hence the strong residual dependence of the resonance lifetimes arises from the interplay of all three particles. Finally, we have performed the calculations for the resonance widths for both 1D and 3D and found qualitatively similar dependence on the mass ratio.

Naturally, the simplicity of our model comes with limitations. We have focused here mainly on the effect of the mass ratio and dimensions on the width of resonances. Certainly, this does not determine the lifetime of resonances alone, as also other factors as $e.g.$ the exact level structure of the two-body subsystem, and the energy difference of two- and three-body states have an effect, as discussed $e.g.$ in Ref.~\cite{nielsen2002}. We have studied this to some extent by analysing three resonances of different energies. Nevertheless, for all three states we have found a strong dependence on the mass ratio over several orders of magnitude, and particular points where the widths even vanish, which can outweight the influence from other factors. Since the effect is so strong, our result can be observed for many kinds of systems in which there is an energy spectrum of three-body resonances and dimer-particle continua as discussed in our article. Moreover, the intriguing oscillatory behaviour indicates that few-body systems can be candidates for the existence of bound states in the continuum.

At the moment we cannot provide a simple explanation for our results, therefore a more thorough analysis is required to bring out the underlying mechanism for stabilization. An approach using a single-particle picture in an effective Born-Oppenheimer or hyperspherical potential curve might be a promising path to follow.

\bmhead{Acknowledgments} We thank M.~A.~Efremov for fruitful discussions. L.~H. is supported by the RIKEN special postdoctoral researcher program. P.~N. acknowledges support from the JSPS Grants-in-Aid for Scientific Research on Innovative Areas (No. JP23K03292).

\bibliography{MassRatioResonances4}

\end{document}